\documentstyle[prl,aps,twocolumn,epsf]{revtex}
\newcommand{\be}{\begin{equation}}
\newcommand{\ee}{\end{equation}}
\newcommand{\bea}{\begin{eqnarray}}
\newcommand{\eea}{\end{eqnarray}}

\begin{document} 

%\twocolumn[\hsize\textwidth\columnwidth\hsize\csname %
%@twocolumnfalse\endcsname

\title {A relation between the resonance neutron peak and ARPES data in
cuprates}
\author{Ar. Abanov and Andrey V. Chubukov}
\address{
Department of Physics, University of Wisconsin, Madison, WI 53706}
\date{\today}
\draft
\maketitle 
\begin{abstract}

We argue that the resonant peak observed in neutron scattering experiments on
superconducting cuprates and the peak/dip/hump features observed in ARPES
measurements are byproducts of the same physical phenomenon. We 
argue that both
are due to feedback effects on the damping of spin
fluctuations in a $d-$wave superconductor. 
We consider the spin-fermion model at strong coupling, 
solve a set of coupled integral equations for fermionic and bosonic propagators
and show that the dynamical spin
susceptibility below $T_c$ possesses the resonance peak at $\Omega_{res}
\propto \xi^{-1}$.
 The scattering of these magnetic excitations
 by electrons gives rise to a 
 peak/dip/hump behavior 
of the electronic spectral function, the peak-dip separation is 
exactly $\Omega_{res}$. 
\end{abstract}

%]
\narrowtext

One of the most intriguing recent developments in the physics of high $T_c$
materials is the realization that not only the normal but also the 
superconducting
state of cuprates is not described by a weak coupling theory.
In particular, ARPES experiments on Bi$2212$ 
have demonstrated~\cite{norman,shennat} 
 that even in slightly overdoped cuprates at $T \ll T_c$, 
the spectral function $A({\bf k},\omega)$  
near $(0,\pi)$ does not possess a 
single quasiparticle peak at $\omega = \sqrt{\Delta^2_{\bf k} + \epsilon^2_{\bf k}}$, 
where $\Delta_{\bf k}$ is
 the superconducting gap  and $\epsilon_{\bf k}$ is the fermionic dispersion.
Instead, it displays
 a sharp peak which virtually does not disperse with $k$, a dip
at frequencies right above the peak, and then a broad maximum (hump) 
which disperses with $k$ and gradually recovers  
the normal state dispersion~\cite{norman}.
Simultaneously, the neutron scattering experiments on near 
optimally doped $YBCO$~\cite{neutrons} and $Bi2212$~\cite{neutrons2}
 at $T \ll T_c$
have detected
a sharp resonance peak in the dynamical structure factor $S(q,\Omega) \propto 
\chi^{\prime \prime} (q,\Omega)$ centered
at $q=Q = (\pi,\pi)$ and at frequencies $\sim 40$ meV. 

In this communication we show that the resonance peak in $S(Q,\Omega)$ 
and the peak/dip/hump features in $A(k,\omega)$ can be 
explained simultaneously by 
strong interaction between electrons and their collective spin degrees of
freedom which near the 
antiferromagnetic instability are peaked at or near $Q$. 
 Specifically, we demonstrate that a $d$-wave superconductor
possesses {\it propagating}  collective spin excitations 
at frequencies smaller than twice  the maximum value of
the $d-$wave gap. The propagating spin modes 
give rise to a sharp peak in $S(Q,\Omega)$ at $\Omega = \Omega_{res} \propto 
\xi^{-1}$ where $\xi$ is the spin correlation
length. The 
interaction with collective spin excitations yields the fermionic self-energy
$\Sigma_\omega$ which at $T=0$ has no imaginary part 
up to a frequency $\omega_0$ which exceeds the measured superconducting 
gap by exactly $\Omega_{res}$.

The point of departure for our analysis is the spin-fermion model for
cuprates which is argued~\cite{chubukov} to be the low-energy theory for
Hubbard-type lattice fermion models. The model is described by
\begin{eqnarray}
{\cal H} &=&
  \sum_{{\bf k},\alpha} {\bf v_F} ({\bf k}-{\bf k}_F)  
 c^{\dagger}_{{\bf k},\alpha} c_{{\bf k},\alpha}
+ \sum_q \chi_0^{-1} ({\bf q}) {\bf S}_{\bf q} {\bf S}_{-{\bf q}} +\nonumber \\
&&g \sum_{{\bf q,k},\alpha,\beta}~
c^{\dagger}_{{\bf k+ q}, \alpha}\,
{\bf \sigma}_{\alpha,\beta}\, c_{{\bf k},\beta} \cdot {\bf S}_{\bf -q}\, .
\label{intham}
\end{eqnarray}
Here  $c^{\dagger}_{{\bf k}, \alpha} $ is the fermionic creation operator
for an electron with crystal momentum ${\bf k}$ and spin $\alpha$,
$\sigma_i$ are the Pauli matrices,  and 
 $g$  is the coupling constant which measures the strength of the 
interaction between fermions and the collective bosonic spin
degrees of freedom. The latter are described  by 
 ${\bf S}_{\bf q}$ and are characterized by a bare  spin susceptibility 
$\chi_0 ({\bf q}) = \chi_0 \xi^2/(1 + (q-Q)^2 \xi^2)$.

Eq. (\ref{intham}) gives rise to fermionic and bosonic self-energies and 
is particularly relevant for fermions near
hot spots -- the 
points at the Fermi surface separated by $Q$. In cuprates, the hot spots are
located near $(0,\pi)$ and symmetry related points. 
The presence of hot spots is essential for our consideration because
the fermions near these points are mostly affected
by the interaction with antiferromagnetic spin fluctuations, and
at the same time, they produce
 the dynamical part of the spin propagator
because a spin fluctuation with a momentum near $Q$ can only decay into
fermions near hot spots.

The normal state properties of the spin-fermion model have recently 
been analysed and
compared with the experiments~\cite{chubukov,chub-morr}. 
It was argued that the experimental situation in cuprates
corresponds to a strong coupling limit $R = {\bar g}/v_F \xi^{-1} 
\gg 1$, where ${\bar g} = g^2 \chi_0$ is the measurable effective
 coupling constant.
The clearest experimental indication for this is the absence of the 
sharp quasiparticle peak in the normal state ARPES data for optimally doped and
and underdoped cuprates~\cite{norman,shennat}.
At strong coupling, a conventional perturbation theory does not work, but it
turns out that one can single out the most divergent diagrams, and 
incorporate them
into a new ``mean-field" ground state. This ``mean-field" theory becomes exact
if one formally sets the number of hot spots
$N =8$ to infinity. At finite $N$, the expansion around the new vacuum holds in
in $(1/N) \log R$, but the prefactors are very small~\cite{chubukov}
 such that for practical purposes, one can
restrict with the $N=\infty$ theory except very near the
antiferromagnetic transition.

This $N=\infty$ theory 
has some similarities with the mean-field $d=\infty$ theories
~\cite{kotliar}: it incorporates the
dominant ($\sim R$) self-energy correction which depends only on
frequency, 
and also includes the dominant bosonic self-energy which is 
the spin polarization bubble made of the renormalized fermions.
The corresponding set of
self-consistent 
equations is presented in Eq.~(\ref{set}) for a superconducting state. The normal
state results are obtained by setting $\Delta =0$. 

The key physical effect which the $N=\infty$ theory
describes is the progressive desctruction, with
increasing $R$, of the coherent quasiparticle peak.
 At the same time, the fermionic
incoherence has no feedback on 
spin susceptibility which still has a simple
relaxational form: 
$ \chi^{-1} (q,\Omega) = 
\chi_0 \xi^2/(1 + (q-Q)^2 \xi^2 - i \Omega/\omega_{sf})$ 
where $\omega_{sf} =
(2\pi/N)~(v_F \xi^{-1})/R$~\cite{chubukov}.
The absence of the feedback effect on spins is a quite general
consequence of the fact that 
fermionic self-energy, abeit strong, has no dependence on the quasiparticle
momentum~\cite{kadanoff}. 

In the superconducting state, this argument does not apply any more because
superconducting and normal state Green's functions have different momentum
dependences
% which can be interpreted as coming from a momentum-dependent
%superconducting self-energy. 
As a result, the feedback effect on spins
is present, and already at the $N=\infty$ level
one has to solve a set of coupled integral equations 
for the fermionic propagator and the spin polarization
operator.
This is the key intent of the present work. We however
will not attempt to self-consistently find also the 
pairing susceptibility which in the spin-fermion model 
results from multiple spin-fluctuation exchanges in the particle-particle
channel~\cite{pairing}. Instead, we assume that below $T_c$
 the pairing susceptibility 
is a conventional $\delta$-function of a total momentum and
frequency of a pair with the $d-$wave amplitude $\Delta^2_k$.
In other words, we will not distinguish between the true superconducting gap
and the pseudogap. The full consideration should indeed include pairing
fluctuations into the self-consistent procedure. 
We will also 
%assume 
%that the fermionic and bosonic self-energies are dominated by the
%regions near hot spots where $\Delta_k \approx \Delta$
%i.e., will 
neglect the processes
which scatter fermions near $(0,\pi)$ into fermions with momenta along zone
diagonal where the $d-$wave gap is absent. The contributions from 
these processes
soften sharp features associated with the $k-$independent gap, but are likely
to be small numerically as they
involve high energy spin fluctuations  with momenta far from $Q$.
Still, however, we will fully explore
the fact that for $d_{x^2 - y^2}$ symmetry of the gap, 
$\Delta_{k+Q} = - \Delta_{k}$.

A simple experimentation
shows that for a $\delta-$functional pairing susceptibility,
 the $N=\infty$ theory 
in the
superconducting state has the same self-consistent structure as in the normal
state but operates with normal and anomalous
fermionic Green's functions 
 $G(k, \omega) = G_0 (k, \omega)/(1 + \Delta^2_k ~G_0 (k, \omega)~G_0 (-k, -
\omega))$ and $F(k, \omega) = -i 
\Delta_k/(1 + \Delta^2_k ~G_0 (k, \omega)~G_0 (-k, -
\omega))$  where 
$G^{-1}_0 (k,\omega) = \omega - \Sigma_{\omega} - {\bf v}_F ({\bf k} - {\bf
k}_F)$ is the zero-order Green's function for $1/N$
expansion. This $G_0 (k,\omega)$ contains a self-energy which results from an 
exchange of a spin fluctuation with $ \chi (q,\Omega) = 
\chi_0 \xi^2/(1 + (q-Q)^2 \xi^2 - \Pi_{\Omega})$ 
where the spin polarization bubble $\Pi_{\Omega}$  
is by itself a convolution of $GG$ and $FF$.
 This construction
yields a set of two coupled integral equations
\begin{eqnarray}
\Sigma_{\omega} &=&3 i g^2 \int \frac{d^2 q d\Omega}{(2\pi)^3}~  
G(k +q, \omega + \Omega)~\chi (q,\Omega)\nonumber \\
\Pi_\Omega &=&-2Ni{\bar g} \xi^2~\int \frac{d^2 k d\omega}{(2\pi)^3}~  
(G(k,\omega)~G(k+Q, \omega + \Omega) \nonumber \\
&& + F(k,\omega)~F(k+Q,\omega))
\label{set} 
\end{eqnarray}
These equations has to be solved to leading order in $1/N$
(this, e.g., eliminates $k-$dependence in $\Sigma$). Higer order self-energy
and vertex correcions are small in $1/N$ and we neglect them.     

It is instructive to consider first the solution of Eqs (\ref{set}) in the
weak coupling limit $\Delta \ll {\bar g} \ll v_F \xi^{-1}$. To first
approximation, $\Pi_\Omega$ can then be evaluated with the free fermion
Green's functions. This has been done before~\cite{mazin}
 and we just quote the result: in the superconducting state,
$\Pi_\Omega$ has both real and imaginary parts. 
The $Im \Pi_\Omega =0$ for $\Omega < 2 \Delta$, it 
jumps at $\Omega = 2\Delta$ 
to $\pi \Delta/\omega_{sf}$, 
and then increases and approaches $\Omega/\omega_{sf}$ at $\Omega \gg 2
\Delta$.  This behavior is similar
to that in an $s-$wave superconductor except for the jump
which is absent in $s-$wave case  and is directly related to the fact that
$\Delta_{k+Q} = - \Delta_k$. 

 The  $Re \Pi_\Omega$ can be obtained either directly or using the
Kramers-Kronig relation. At $\Omega \ll \Delta$, we
 have $Re \Pi (\omega) = (\pi/8)~\Omega^2/(\Delta \omega_{sf})$. 
It diverges at $2\Delta$ 
as $\Pi_\Omega = (\Delta/\omega_{sf}) \log (2\Delta/|2\Delta
- \Omega|)$ because of the jump in $Im \Pi_\Omega$,
and decreases at larger frequencies. Due to the divergence,
$Re \Pi_\Omega$ reaches 1 at a frequency $\Omega_{res}$ which
is less than $2\Delta$, i.e., 
when $Im \Pi_\Omega$ is still zero. Explicitly, 
$\Omega_{res} = 2\Delta (1 -Z)$ where
$Z \propto e^{-\omega_{sf}/(2\Delta)}$. Near $\Omega_{res}$,
 $\chi(Q,\Omega) \propto Z /(\Omega - \Omega_{res} - i \delta)$, i.e., the
dynamical structure factor has a resonance peak.

Consider next the fermionic spectral function.
 Without self-energy corrections,
$A(k, \omega)$ near a Fermi surface resonates at $\omega_{res} = \Delta$. The
self-energy gives rise to a fermionic decay. 
For an $s-$wave superconductor, the onset frequency for a decay
is $3\Delta$, and $Im \Sigma (k,\omega)$ emerges as $(\omega -3
\Delta)^{1/2}$~\cite{varma}. The presence of the resonance mode qualitatively 
changes this picture because a fermion can decay into this 
mode starting from $\omega_0 < 3\Delta$.
 A simple power counting shows that this 
process yields a finite jump of
$Im \Sigma_\omega$ at the onset frequency
and hence the logarithmical
singularity in $Re \Sigma_\omega$. The latter in turn increases the
self-energy at $\omega \sim \Delta$ 
 and shifts downwards $\omega_{res}$ (which is the measured gap),
and the onset 
frequency for $Im \Pi_\Omega$ which, as one can easily demonstrate,
 exactly equals $2\omega_{res}$. 
The amounts of the 
shifts and the amplitude of the jump in $Im \Sigma_\omega$ can be 
obtained explicitly from
 Eqs. (\ref{set}). We found $\omega_0 = 3\Delta (1 - \epsilon)$,
$\omega_{res} = \Delta (1- \epsilon)$, and 
$\delta (Im \Sigma_{\omega_0}) = (\pi \Delta/\log 2)~ \epsilon$, 
where $\epsilon =(3 \log 2 \sqrt{N}/(64 \sqrt{\pi})~
({\bar g}/\Delta)^{1/2}~e^{- \omega_{sf}/(2\Delta)}$. 

\begin{figure} [t]
\begin{center}
\leavevmode
\epsfxsize=3.0in 
\epsfysize=2.6in 
\epsffile{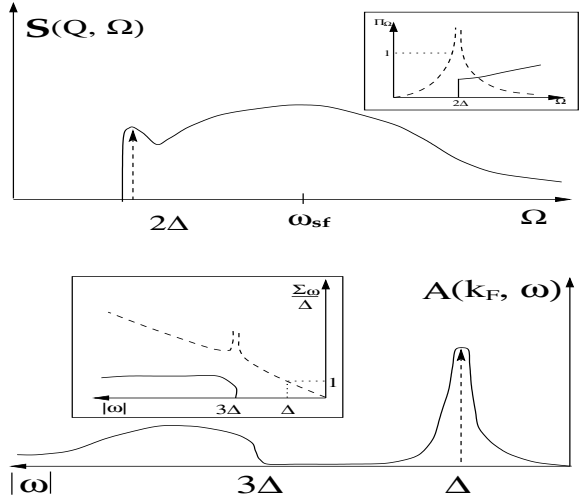}
\end{center}
\caption{
The $T=0$ weak coupling behavior of the dynamical structure factor and
the spectral function. Solid lines are schematic 
solutions of Eqs. (\protect\ref{set}) broadened by
experimental resolution. Without resolution, the peaks are $\delta$-functions
as indicated by arrows.
The insets show the
spin polarization operator $\Pi_\Omega$ and the fermionic self-energy
$\Sigma_\omega$ (solid lines -- imaginary parts, dashed lines -- real parts).
The vertical dashed lines denote logarithmical singularities.}
\label{fig1}      
\end{figure}

We see 
that the $d-$wave form of the gap yields 
qualitative changes in the
system behavior compared to the 
$s-$wave case, 
%(the reduction of $\omega_{res}$ and the onset
%frequencies for fermionic and bosonic damping), 
but at small coupling 
%($\Delta \ll \omega_{sf}$), 
these changes are exponentially small and can hardly be measured. 
In particular, the resonance  peak in $S(Q,\omega)$ 
should be smeared out already by a small experimental resolution.
 The weak coupling 
results are
%behavior of $S(Q,\Omega)$ and $A(k,\omega)$ is 
shown in Fig~\ref{fig1}.

We now turn to strong coupling limit 
${\bar g} \gg \Delta \gg (\omega_{sf}~{\bar g})^{1/2}$.
Solving Eq. (\ref{set}) as explained below, we found
that the peak frequency
in $A(k_F, \omega)$ (i.e., the measured gap) 
is now $\omega_{res} = {\bar \Delta}
\sim \Delta^2/{\bar g}$. 
The nonzero $Im \Pi_\Omega$ and 
$Im \Sigma_\omega$ appear respectively 
at $2 {\bar \Delta}$ and 
$\omega_0  = {\bar \Delta} (1 + a)$ where $a \propto 
(\omega_{sf}/{\bar \Delta})^{1/2} \propto \xi^{-1}$. The amounts of jumps 
in $Im \Pi_\Omega$ and $Im \Sigma_\omega$ both
 scale as $a^{1/2}$
and completely disappear at $\xi = \infty$ when 
$\omega_{sf} =0$. Above the threshold, $Im \Sigma_\omega$ first increases 
as $Im \Sigma_\omega
\propto (\omega - \omega_0)^{\nu}$ where $\nu = {\sqrt 3} -1$, and then
recovers the normal state, $\sqrt{\omega}$ behavior. Substituting
this $\Sigma$ into $G (\omega)$, we found that it possesses a peak at
$\omega_{res}$, a dip at $\omega_0$ and a 
hump at $\omega_{hump} = \omega_{res}(1 + b)$ where $ b \sim 
({\bar g}/{\bar \Delta})^{3/(4\nu)}~R^{1/(2\nu)} \log R \propto
\xi^{-1/(2\nu)} \log \xi$.
At $\xi = \infty$, peak, dip and hump positions coincide with each other, 
and the peak/dip/hump structures transformes into the  edge singularty: 
$A (\omega) \propto (\omega - \omega_{res})^{-\nu}$.

Further, the fact
that $Im \Pi_\Omega =0$ up to $2 \omega_{res}$ implies, 
via Kramers-Kronig relation, 
that at small frequencies 
$Re \Pi_\Omega \propto \Omega^2/( \omega_{sf} {\bar \Delta})$.
Substituting this result into  $S (q,\omega)$, 
we find that it 
possesses a resonance peak at $\Omega_{res} \sim 
(\omega_{sf} {\bar \Delta})^{1/2} \sim \xi^{-1}
 \ll 2{\bar \Delta}$. 
At $q \neq Q$, the peak disperses
with $q$  
as $\Omega^2 = \omega_{res}^2 ((1 + ((q-Q)\xi)^2$,
just as a conventional spin wave,  untill $\Omega$ reaches 
$2 \omega_{res}$, and at larger frequencies disappears due to damping.  

\begin{figure} [t]
\begin{center}
\leavevmode
\epsfxsize=3.0in 
\epsfysize=2.6in 
\epsffile{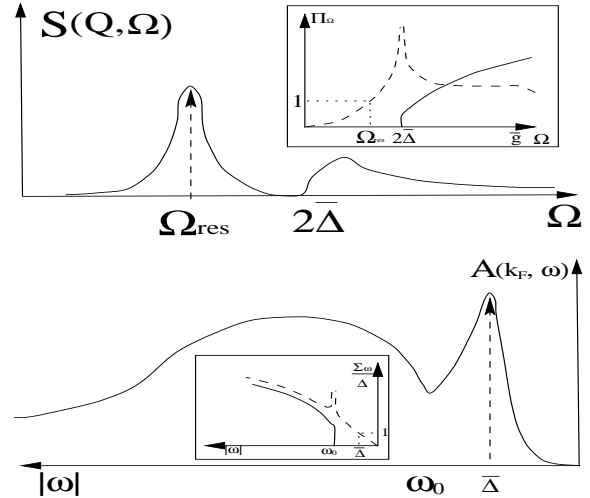}
\end{center}
\caption{
 Same as in Fig~\protect\ref{fig1} but at strong coupling. 
The resonance and onset frequencies are presented in the text.
The spin resonance frequency $\Omega_{res} \propto \xi^{-1}$,
is equal to the distance between
the measured gap ${\bar \Delta}$ and the dip frequency $\omega_0$.
 The hump frequency differes from 
${\bar \Delta}$ roughly by $\xi^{0.7}$.}
\label{fig2}      
\end{figure}

The strong-coupling behavior of $S(Q,\Omega)$ and $A(k,\omega)$
is presented in Fig.~\ref{fig2}. 
We see that (i) $S(q,\omega)$ possesses a sharp resonance peak at $\Omega_{res}
\sim \xi^{-1}$ which
shifts with underdoping  to
lower frequencies, and 
(ii) $A(k_F, \omega)$ possesses a quasiparticle peak at 
$\omega = \omega_{res}$, a dip at $\omega_0 = \omega_{res} + \Omega_{res}$,
 where $Im \Sigma_\omega$ first appears,
and a broad maxumum at a somewhat higher frequency $\omega_{hump}$.
 As the momentum moves away from the Fermi surface, the  spectral function
for frequencies larger than $\omega_0$ disperses
with $k$ and recovers the normal state, non-Fermi liquid
 form with a broad maximum 
at $\omega \sim \epsilon^2_k/{\bar g}$. The quasiparticle peak however
cannot move further than $\omega_0$ because of a strong fermionic damping above
threshold.  We found that it gets pinned at $\omega_0$
and just gradually looses its strength with increasing $k-k_F$.  

We emphasize that although the resonance frequency 
in $S(Q,\omega)$ continuously evolves from weak to strong coupling,
the physics changes qualitatively
between the two limits. At weak coupling, the peak is solely due to 
a jump in $Im \Pi_\Omega$. At 
strong coupling, the jump is almost gone, and the existence of peak is
due to $\Omega^2$ behavior of $Re \Pi_\Omega$ which is related
to vanishing $Im \Pi_\Omega$ below $2 \omega_{res}$. 

We now briefly discuss how we obtained these results. 
 We first integrated over momentum in (\ref{set}) and 
for $R = {\bar g}/(v_F \xi^{-1}) \gg 1$ obtained
%neglecting  $\omega$ compared to $\Sigma_\omega$ in $G_0 (k, \omega)$.
\begin{equation}
\Sigma_\omega = \frac{3R}{8 \pi}~\int d \Omega~
\frac{\Sigma_{\omega +\Omega}}{\sqrt{\Sigma ^{2}_{\omega +\Omega}-\Delta ^{2}}}
\frac{1}{\sqrt{1-\Pi_\Omega }} 
\label{set2a}
\end{equation}
\begin{equation}
\Pi_\Omega= \frac{i}{2}
\int \frac{d \omega}{\omega_{sf}}~\left(
\frac{\Sigma_{\Omega -\omega}~\Sigma_\omega + \Delta ^{2}}
{\sqrt{\Sigma^{2}_{\Omega -\omega}-\Delta ^{2}}~
\sqrt{\Sigma^{2}_\omega-\Delta ^{2}}}+1 \right)
\label{set2b}
\end{equation}
We then performed self-consistent calculations: 
we first obtained regular terms in $Re \Sigma$ and $Re \Pi$ by 
expanding (\ref{set2a}, \ref{set2b}) 
in the external frequencies. We obtained
$Re \Sigma_\omega \propto \omega ({\bar g}/{\bar \Delta})^{1/2}$ 
 and $Re \Pi_\Omega \propto \Omega^2/({\bar \Delta} \omega_{sf})$. 
In both cases, the integrals
are confined to frequencies $\sim {\bar \Delta}$ where the system interpolates
between normal state and superconducting behavior, and 
for estimates, one
can use in the integrands the known normal 
state results for $\Sigma_\omega$ and $\Pi_\omega$.
We then assumed that at some finite frequency $\omega_0$
$Im \Sigma_\omega$ jumps from $0$ to  some finite value, considered the 
onset frequency and the amount of the jump as input parameters, and 
 used Kramers-Kronig relation to
calculate the logarithmicaly singular term in $Re \Sigma_\omega$. Adding
it to a regular $Re \Sigma_\omega \propto \omega$, we find $\omega_{res}$ where
$\omega + Re \Sigma_{\omega_{res}} = \Delta$. 
Substituting next $Re \Sigma_{\omega}$ 
into (\ref{set2b}) and using the spectral
representation for $Im \Pi_\Omega$, we 
find the threshold frequency for $Im \Pi_\Omega$ at $2 \Omega_{res}$ 
and the amount of the jump at the threshold. We
then again use Kramers-Kronig relation to calculate a logarithmically 
singular contribution to $Re \Pi$,
 add it to a regular $Re \Pi_\Omega \propto \Omega^2$, and
subsitute the result into (\ref{set2a}) for $\Sigma_\omega$.
Using again the spectral representation for $Im \Sigma_\omega$,
we find two self-consistent equations for threshold frequency $\omega_0$
and for the amount of the jump at the threshold. 
 These calculations are not exact because we
only know regular $Re \Sigma_\omega$ and $Re \Pi_\Omega$ 
upto  overall numerical factors,
but they yield correct functional forms of the resonance and threshold
frequencies.

We now compare our results with the data. Qualitatively, the peak/dip/hump
behavior of $A(k,\omega)$, the absence of the peak dispersion with $k$,
 and the presence of the dispersing 
resonance mode in $S(q,\omega)$ all agree with the  ARPES and neutron
measurements in $YBCO$ and 
$Bi2212$~\cite{norman,shennat,neutrons,neutrons2}. More
quantitatively, we predict that the peak-dip separation in $A(k,\omega)$ at
a hot spot exactly equals to the resonance frequency in $S(Q,\Omega)$. 
Experimentally, in near optimally doped, $T_c = 87K$
 $Bi2212$, $\omega_0 - \omega_{res}
\approx 42 meV$~\cite{norman}. Recent neutron scattering data~\cite{neutrons2}
 on $Bi2212$ with nearly the same $T_c = 91K$ yielded 
$\Omega_{res} =43 meV$, in
full agreement with the theory. 
We also predict that the peak, dip and hump frequencies should tend to the same
value as $\xi \rightarrow \infty$. At large but finite $\xi$, 
 we predict that 
$(\omega_{hump}/\omega_{res} -1) \propto (\omega_0/\omega_{res} -1)^{\beta}$
where $\beta = (\sqrt{3} + 1)/4 \sim 0.7$.  

We now connect our work to earlier studies. 
That the interaction
with a nearly resonant collective mode peaked at $Q$ 
explains the ARPES data has been
known for some time, and qualitative arguments have been 
displayed first in~
\cite{schr-shen} and then in ~\cite{norman}. Ref~\cite{norman} also conjectured that 
the peak-dip separation may be related to a neutron peak frequency. 
It has been also realized earlier that 
in a $d-$ wave superconducting Fermi gas, $S(Q,\Omega)$
contains a resonance peak exponentially close to
 $2\Delta$~\cite{mazin}. From this perspective, 
the key intension of this work was to
present the quantitative results for cuprates by performing actual strong
coupling calculations, and to explicitly relate ARPES and neutron scattering
data. 

Morr and Pines~\cite{morr}
obtained the spin-wave like dispersion
 in $\chi (q,\Omega)$ below $T_c$ by phenomenologically 
adding the $\Omega^2$ term to the bare susceptibility. 
This term  should be in the form $\Omega^2/\epsilon_F$ as it can
only come from fermions located far away from the Fermi
surface. We have
demonstrated that  at $T < T_c$, 
the spin-fermion model of Eq (\ref{intham})
by itself generates an  $\Omega^2/\Delta$ term which for $\Delta \ll
\epsilon_F$ completely
overshadows a possible bare $\Omega^2$ term.

Morr and one of us~\cite{chub-morr}
considered an approximate
solution of Eq. (\ref{set2b}) assuming that $\Pi_\Omega$ still has
the same purely relaxational form  $i \Omega/\omega_{sf}$
as in the normal state, but $\omega_{sf}$ is momentum dependent.
% This aproximation does not satisfy the 
%Kramers-Kronig relation, but was argued in 
%\cite{chub-morr} to reasonably describe the electronic spectral function
%at finite temperatures.
Comparing our results with ~\cite{chub-morr}, we found that the approximate
solution for $A(k,\omega)$ captures the main features 
of the full solution, i.e., the
peak/dip/hump structure of $A(k,\omega)$, but
yields incorrect values
of the peak and dip frequencies for ${\bar \Delta} \gg \omega_{sf}$. 

J. Brinckman and P. Lee studied the evolution of the resonance peak within the
slave boson theory~\cite{lee}. Their philosophy and the 
results are similar to ours.

To summarize,  we considered the 
superconducting phase of cuprates
and demonstrated that the resonance peak in the
dynamic structure factor and the peak/dip/hump structure of the electronic
spectral function near $(0,\pi)$ can simultaneously be explained 
by a strong spin-fermion interaction. The 
peak-dip separation at a hot spot exactly equals to the resonance neutron
frequency and vanishes at $\xi = \infty$. The peak-hump separation also
vanishes, but with a smaller power of $\xi^{-1}$.

It is our pleasure to thank G. Blumberg, A. Finkelstein,
R. Joynt, G. Kotliar, A. Millis, D. Morr, M. Norman, D. Pines and J. Schmalian 
for useful conversations. 
The research was supported by NSF DMR-9629839.

\end{document}